\documentclass{article}

\usepackage[utf8]{inputenc}
\usepackage{amsmath, amsthm, amssymb, amsfonts, dsfont, mathrsfs, dirtytalk, tikz-cd, adjustbox, url, nicefrac, booktabs,array, graphicx,todonotes,longtable,tabularx, tabu,bm,comment,bm,titlesec,hyperref,enumitem,csquotes}
\usepackage[T1]{fontenc}
\usepackage[english]{babel}
\usepackage[margin=0.7in]{geometry}
\usepackage[backend=biber,style=apa,sorting=nyt,giveninits=true,doi=false,isbn=false,url=false,date=year]{biblatex}
\usepackage[font=small]{caption}
\AtEveryBibitem{\clearfield{month}\clearfield{issn}}
\usepackage{appendix}
\usepackage[capitalise]{cleveref}

\titlespacing*{\paragraph}{0pt}{1ex}{1ex}

\newcommand{\dkl}{\operatorname{D_{K L}}}

\newtheorem{remark}{\bf Remark}[section]
\newtheorem{proposition}{\bf Proposition}[section]

\newtheorem{example}{\bf Example}[section]

\newtheorem{assumption}{\bf Assumption}[section]

\title{A Rosetta Stone Hypothesis for Neurophenomenology: \\ Mathematical Predictions from Predictive Processing}
\author{Lancelot Da Costa$^{1,2,3}$, Anil K. Seth$^{4,5,6}$, Karl Friston$^{1,3}$,\\ Maxwell J. D. Ramstead$^{3, \dagger}$, Lars Sandved-Smith$^{7,\dagger}$\thanks{Correspondence: \textit{lars.sandvedsmith@gmail.com} $\quad ^\dagger$ Joint senior author.}\\}
\date{\small\it
    $^1$VERSES AI Research Lab, Los Angeles, CA 90016, USA\\
    $^2$Department of Mathematics, Imperial College London, London, SW7 2AZ, UK\\
    $^3$Wellcome Centre for Human Neuroimaging, University College London, London, WC1N 3AR, UK\\
    $^4$Sussex Centre for Consciousness Science, University of Sussex, Brighton, UK\\
    $^5$Department of Informatics, University of Sussex, Brighton, UK\\
    $^6$Canadian Institute for Advanced Research, Program on Brain, Mind and Consciousness, Toronto, Canada\\
    $^7$Monash Centre for Consciousness and Contemplative Studies, Monash University, Australia
}

\begin{document}

\maketitle

\vspace{-10pt}

\begin{abstract}
Consciousness science faces the challenge of bridging first-person experience with third-person empirical measurements. Neurophenomenology aims to build such `generative passages' connecting the content of experience with behavioural and neuroscientific data. However, the mathematical machinery for such bridges remains underdeveloped. Here we develop a Rosetta Stone hypothesis from predictive processing, where beliefs serve as a central hub connecting phenomenology, behaviour, and neural dynamics. This hinges on a central technical assumption that phenomenology is a function of beliefs. We pursue a conditional approach: if this assumption holds, then certain predictions mathematically follow. We derive predictions for subjective similarity judgements, cognitive metabolic cost, subjective cognitive effort, and time perception. We review the connection between beliefs and neural dynamics to complete the generative passage for neurophenomenology, omitting the connection between beliefs and behaviour as this is already well-documented elsewhere. Testing our predictions will inform the validity of the central assumption connecting beliefs and phenomenology, and advance the neurophenomenology research programme.

\end{abstract}


\tableofcontents

\section{Introduction}

\subsection{The neurophenomenology challenge}
This work is situated within the scope of a research program known as neurophenomenology \parencite{varelaMethodologicalRemedyHard1996, varelaNaturalizationPhenomenologyTranscendence1997} and especially its more recent mathematical and computational expressions \parencite[Table~1]{magoInclusiveComputationalNeurophenomenologyInprep}.

Phenomenology concerns the rigorous descriptive study of the various kinds of conscious experience, outlining the essence or necessary properties of each type of conscious experience \parencite{husserl2012ideas1}. Since the 1990s, there has been an interest and coordinated effort to explicitly combine first-person phenomenological methods, generating detailed qualitative descriptions of lived experience, with third-person neuroscientific techniques used to measure and quantify brain activity \parencite{petitot1999naturalizing}.
This program, known as `neurophenomenology', was articulated originally by Varela \parencite{varelaMethodologicalRemedyHard1996, varelaNaturalizationPhenomenologyTranscendence1997}.

What set neurophenomenology apart from extant fields was its emphasis on `generative passages': the explicit mutual constraints and virtuous informative cycles linking first- and third-person methods. Neurophenomenology differs from other approaches to phenomenology in its aim to build such generative passages, with neurobiological data and models constrained by, and constraining, models and data from first-person phenomenological methods, rather than proposing a theory that can distinguish between conscious and non-conscious processing, or proposing a mere isomorphism between first- and third-person descriptions. Mathematical language would offer a kind of ontologically neutral bridge between these two domains; in its original formulation, the mathematics of dynamical systems theory were seen as especially apt as a bridge \parencite{roy1999beyond}.

As Varela wrote:
`A more demanding approach will require that the isomorphic idea is taken one step forward to provide the passage where the mutual constraints not only share logical and epistemic accountability, but they are further required to be \textit{operationally generative}, that is, where there is a \textit{mutual circulation and illumination} between these domains proper to the entire phenomenal domain. This is to say, we must be prepared to be in a position to generate (in a principled manner) reduction analysis [i.e., subjective descriptions of lived experience] and eidetic descriptions [i.e., descriptions of the necessary properties of kinds of conscious experience] that are rooted in an explicit manner to biological emergence' \parencite{varela1997naturalization}, emphasis added.

Despite the detailed conceptual advances made by Varela and colleagues, which partially motivated the successful reintroduction of consciousness as a worthy or non-suspect topic of scientific investigation, the question of how one might \textit{formalise} generative passages in a principled manner remains a hotly debated issue---and a relatively open challenge.

\begin{figure}[ht!]
    \centering
    \includegraphics[width=0.6\textwidth]{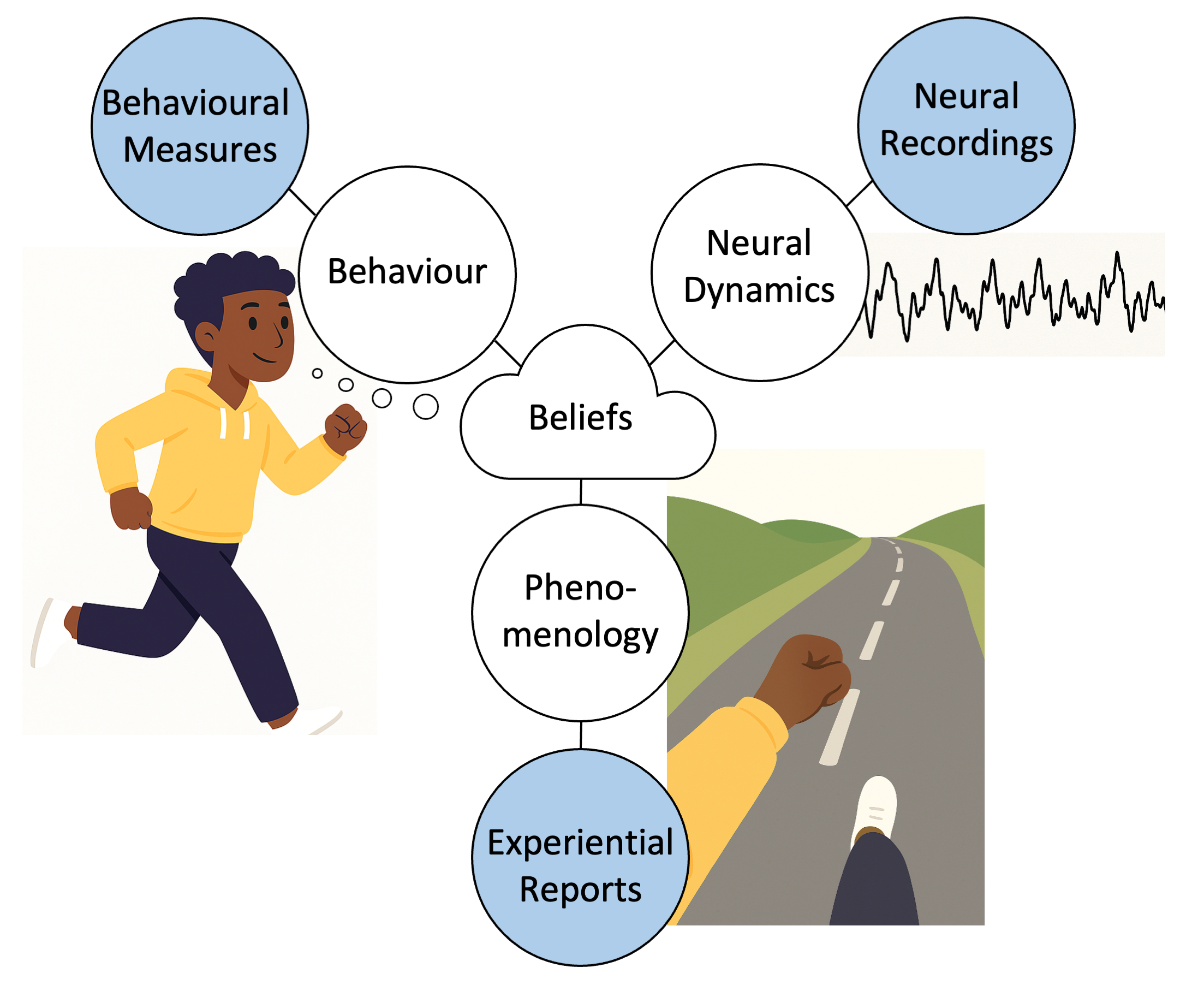}
    \caption{\textbf{A Rosetta Stone for Neurophenomenology.} We posit that beliefs serve as the central hub connecting phenomenology, behaviour, and neural dynamics. Beliefs here are probability distributions: approximate posterior beliefs about the causes of sensations, as used in predictive processing and Bayesian inference. Each connection represents a bridge that can be empirically investigated: phenomenology can be accessed through experiential reports, behaviour through behavioural measures, and neural dynamics through neural recordings. We investigate predictions from this Rosetta stone under the central technical assumption that phenomenology corresponds to beliefs.}
    \label{fig: rosetta stone}
\end{figure}

\subsection{Predictive processing}

Our approach to neurophenomenology builds on the predictive processing premise that cognition can be described as a process of inference about the external causes of sensory input. This lineage is often traced back to Helmholtz's notion of `unconscious inference' in perception, which prefigures modern views of the brain as constructing hypotheses about the world from ambiguous data \parencite{helmholtzHelmholtzTreatisePhysiological1962}. In contemporary neuroscience, this idea was revived in the predictive coding account of cortical hierarchies, in which top--down predictions are compared against bottom--up prediction errors \parencite{raoPredictiveCodingVisual1999,fristonTheoryCorticalResponses2005}. The broader `Bayesian brain' hypothesis then reframed perception and learning as approximate Bayesian inference under uncertainty \parencite{knillBayesianBrainRole2004}. More recent formulations in the free-energy principle and active inference frameworks generalise this inferential view to include action and control, treating perception, learning, and behaviour as different facets of a single imperative: maintaining and refining a generative model by optimising a variational free energy functional (also known in statistics as an evidence lower bound \parencite{bleiVariationalInferenceReview2017,bealVariationalAlgorithmsApproximate2003}) \parencite{fristonFreeenergyPrincipleUnified2010,dacostaActiveInferenceDiscrete2020,parrActiveInferenceFree2022}.

In this work, we consider an organism interacting with its environment. We denote external (environmental) states by $s$, internal states by $\mu$, and sensory states (observations) by $o$. Depending on the level of description, the `organism' could be a whole brain coupled to an external world, a brain region interacting with other regions (as its effective `external' states), or even a single neuron that receives synaptic input and acts on downstream neurons through firing. We assume that internal states can be described as parameterising beliefs $q_\mu(s)$ about external states \parencite{dacostaBayesianMechanicsStationary2021,ramsteadBayesianMechanicsPhysics2022,fristonFreeEnergyPrinciple2023a}. Beliefs are meant in a technical sense as probability distributions---typically approximate posterior beliefs---as commonly used in predictive processing and Bayesian statistics. We will usually assume that these beliefs evolve so as to (approximately) solve a variational inference problem, tracking external states $s$ given sensations $o$ under an implicit generative model $p(o,s)$. Equivalently, beliefs evolve to optimise a variational free energy functional $F$:
\begin{equation}
    \label{eq: fe}
    \mu \mapsto q_\mu(s),\qquad
    \mu \searrow \operatorname{F}\!\left[q_\mu,o\right]
    := \dkl\!\left[q_\mu(s) \mid p(s \mid o)\right]-\log p(o).
\end{equation}
This captures the Bayesian brain hypothesis (under a variational implementation) and, more generally, the core inferential objective underlying the free-energy principle and active inference frameworks. There is a mathematical justification for why internal states of organisms may often be described as encoding beliefs about external states, and for why their dynamics may be cast as (approximate) variational inference---afforded by Bayesian mechanics (see Appendix \ref{app: pp}) \parencite{dacostaBayesianMechanicsStationary2021,ramsteadBayesianMechanicsPhysics2022,fristonFreeEnergyPrinciple2023a}. In what follows, we use belief dynamics as a common mathematical currency for building `generative passages' between first-person experiential reports and third-person measurements of behaviour and neural activity, under the central assumption that phenomenology is a function of beliefs.

\subsection{The Rosetta Stone hypothesis}
In this work we develop the hypothesis that beliefs serve as a central hub connecting phenomenology, behaviour, and neural dynamics (\cref{fig: rosetta stone}). This furnishes a generative passage between subjective experiential reports, objective behavioural measures, and objective neural recordings. This hypothesis hinges on our central technical assumption which states that phenomenology is a function of beliefs. We pursue a conditional approach: \textit{if} the assumption is true \textit{then} there are consequences and testable predictions that follow. These predictions can in turn be used to test this central technical assumption or refine it by illuminating the nature of the phenomenology-belief correspondence. This framework offers a \emph{theory of use} to consciousness science: we use beliefs as a bridging principle to characterise phenomenology and derive testable predictions, without making specific claims about where phenomenological content might lie in an organism's beliefs---which would amount to a theory of consciousness. 

Why is it helpful to consider beliefs as a central hub? Pairwise relations between correlates of phenomenology---reports, behaviour, and neural recordings---can be studied directly. Our claim is that beliefs become especially useful once the aim is to obtain a \emph{mechanistic understanding} of the relationship between phenomenology, behaviour, and neural dynamics within a single explanatory framework, and to derive predictions that generalise across tasks and measurement modalities. In predictive processing, beliefs are model-dependent variables that mediate between sensory evidence, action, and neural processing. Furthermore, they are a natural level at which to formalise quantities such as similarity, precision, confidence, and information length, which are not straightforwardly defined at the level of raw reports or neural recordings alone. In this sense, beliefs are introduced not to replace phenomenology or its correlates, but to provide a common mathematical currency for relating them.

\paragraph{Scope, organisation, and contribution.} In light of the minimal commitments proposed for computational neurophenomenology \parencite[Table~1]{magoInclusiveComputationalNeurophenomenologyInprep}, our approach treats phenomenology as an explanandum via first-person reports (similarity judgements, effort ratings, duration judgements), specifies explicit link hypotheses between phenomenology, beliefs, and neural dynamics, and derives falsifiable predictions, aiming to make the generative passage operational. Concretely, in \cref{sec: 2} we state our central technical assumption that phenomenology is a function of beliefs. In \cref{sec: 3}, we examine the consequences of this correspondence (cf. \cref{fig: rosetta stone}, bottom) both mathematically and empirically, stating predictions for subjective experiential reports. We expose a geometry for phenomenology enabling a precise characterisation of phenomenological differences between subjects (\cref{sec: mathematical}). We then make predictions for (1) subjective similarity judgements (\cref{sec: subjective}), (2) cognitive metabolic cost and subjectively experienced cognitive effort (\cref{sec: cost}), and (3) the experience of temporal duration (\cref{sec: time}). In \cref{sec: 4}, we synthesise the relevant predictive processing literature connecting beliefs and neural dynamics, furnishing a generative passage to neural recordings (cf. \cref{fig: rosetta stone}, top right). We largely set aside the generative passage between beliefs, behaviour, and behavioural measures (cf. \cref{fig: rosetta stone}, top left) since this is already covered extensively in related literature: see \parencite{dacostaActiveInferenceDiscrete2020,smithStepbystepTutorialActive2022,parrActiveInferenceFree2022} for reviews.

\subsection{Related work}

\paragraph{Active inference work on generative passages.}
Most related to our work is \parencite{ramsteadGenerativeModelsGenerative2022, sandved-smithDeepComputationalNeurophenomenology2024, sandved2021towards}, which leverage active inference accounts of predictive processing to model subjective experience. They propose that once a type of phenomenological experience is formalised, that description can be used to constrain candidate models of the neural dynamics that might realise or enable that experience via \emph{generative passages}. \textcite{sandved-smithDeepComputationalNeurophenomenology2024} in particular argues, as we do, that the mathematics of belief dynamics (Bayesian mechanics) is particularly apt to enable generative passages. However, these works do not explicitly develop the mathematical machinery needed to (i) \emph{characterise phenomenology quantitatively} (e.g., via a geometry, distances, or lengths), and (ii) \emph{investigate the form of the mapping} implementing the passage from beliefs (e.g., approximate posteriors) to first-person phenomenology, together with the constraints and predictions such a mapping induces. They focus instead on providing an active inference account of the first-person experience itself, as described through phenomenological methods. See \textcite{sandved2021towards} for a worked example applied to the phenomenology of focused attention. Our contribution builds on their work by making the bridge mathematically explicit and deriving portable quantities that can be carried across tasks and linked to behavioural and neural correlates.

\paragraph{Predictive processing theories of consciousness.}
In contrast to our theory of use, other work makes specific claims about where phenomenological content might lie in an organism's beliefs, which amounts to a proper theory of consciousness \parencite{ramsteadInnerScreenModel2023,sandved-smithMetacognitiveParticlesMental2024,WHYTE20264,laukkonenBeautifulLoopActive2024}.

\paragraph{Neural-network simulations of visual phenomenology.}
Other approaches use neural network models to simulate specific forms of visual phenomenology. For example, Suzuki and colleagues illustrated this by adapting a deep convolutional neural network (AlexNet \parencite{krizhevskyImageNetClassificationDeep2012}) to simulate the phenomenology of visual hallucinations \parencite{suzukiDeepDreamVirtualReality2017}. This was recently extended using coupled discriminative and generative networks to target distinct hallucination profiles associated with different aetiologies \parencite{suzukiModellingPhenomenologicalDifferences2024}. While inspired by Bayesian perspectives on perception, this line of work does not investigate the consequences of a correspondence between beliefs and first-person experiential reports.

\section{Central Technical Assumption}
\label{sec: 2}

We explore the mathematical implications of a central hypothesis: that phenomenological content corresponds to beliefs. We pursue a conditional approach: \emph{if} the hypothesis holds, \emph{then} these are the consequences. This allows us to derive testable predictions about phenomenology without claiming to resolve fundamental questions about the nature of consciousness. The predictions we derive can, in turn, be used to empirically test or refine this central assumption, advancing the computational neurophenomenology research programme.

\begin{assumption}[Central Technical Assumption]
    \label{as: core}
    We adopt the hypothesis that phenomenological content corresponds to beliefs. Let $p \in \mathcal{P}$ denote phenomenological content and let $\varphi \colon \mathcal{Q} \to \mathcal{P}$ be a mapping from beliefs to phenomenology. We assume that phenomenology is a function of beliefs
    \begin{equation}
        p = \varphi(q_\mu),
    \end{equation}
    where $q_\mu$ is the approximate posterior belief encoded by the internal states $\mu$ of the system. In particular, this makes $\varphi$ a surjective map onto phenomenology.
\end{assumption}

\begin{example}
    \label{eg: nature}
    We can consider four nested possibilities about the nature of $\varphi$, organised by decreasing strength:
    \begin{enumerate}[noitemsep]
        \item \emph{Identity.} $\varphi$ is the identity map---all beliefs are phenomenological, a position consistent with the view that consciousness is widespread in nature.
        \item \emph{Marginalisation.} $\varphi$ is a marginalisation onto a subset of beliefs that have phenomenological content. E.g., in the case of the brain, this postulates that phenomenology relates to beliefs encoded by particular brain systems.
        \item \emph{Pushforward.} $\mathcal{P}$ is a space of probability distributions and $\varphi$ arises from a map between the underlying sample spaces of $\mathcal{Q}$ and $\mathcal{P}$---for instance, a coarse-graining that groups fine-grained beliefs into coarser categories (see \cref{app: induced fisher} for details).
        \item \emph{Arbitrary.} $\varphi$ is an arbitrary deterministic function, where $\mathcal{P}$ need not be a space of probability distributions.
    \end{enumerate}
    Under interpretations 1--3, $\mathcal{P}$ is a space of probability distributions, while for interpretation 4, $\mathcal{P}$ could be arbitrary. Our framework applies under any of these interpretations, but the strength of the resulting predictions differs. Table~\ref{tab: phi-predictions} summarises how the article's main predictions depend on the form of $\varphi$.
\end{example}

\begin{table}[h!]
\centering
\footnotesize
\setlength{\tabcolsep}{2pt}
\renewcommand{\arraystretch}{0.96}
\caption{\textbf{Predictions under different belief--phenomenology mappings.} Each entry gives the predicted pattern and headers point to the section where each prediction is developed.}
\label{tab: phi-predictions}
\begin{tabularx}{\textwidth}{@{}>{\raggedright\arraybackslash}p{0.12\textwidth} *{4}{>{\raggedright\arraybackslash}X}@{}}
\toprule
Mapping $\varphi$ & Similarity judgements (\cref{sec: subjective}) & Cognitive metabolic costs (\cref{sec: cost}) & Subjective cognitive effort (\cref{sec: cost}) & Time perception (\cref{sec: time}) \\
\midrule
Identity & If similarity judgements track phenomenological distances: attended/high-confidence percepts are judged more distinct. & Larger belief length predicts larger neural entropy production and brain metabolic expenditure. & Subjective cognitive effort tracks phenomenological length. & If belief or phenomenological length tracks subjective time: attended/high-confidence sequences feel longer. \\
Marginalisation & Identity $\varphi$ prediction holds when precision affects phenomenological subset. & Same as identity $\varphi$. & Same as identity $\varphi$. & If belief length tracks time: same as identity $\varphi$. If phenomenological length tracks time: prediction holds when precision affects phenomenological subset. \\
Pushforward & Identity $\varphi$ prediction is preserved only when $\varphi$ preserves some precision differences. & Same as identity $\varphi$. & Same as identity $\varphi$. & If belief length tracks time: same as identity $\varphi$. If phenomenological length tracks time: prediction is preserved only when $\varphi$ preserves some precision differences. \\
Arbitrary & No general prediction without extra assumptions on $\varphi$. & Same as identity $\varphi$. & Same as identity $\varphi$. & If belief length tracks time: same as identity $\varphi$. If phenomenological length tracks time: no general prediction without extra assumptions on $\varphi$. \\
\bottomrule
\end{tabularx}
\end{table}

\section{Phenomenology and Beliefs}
\label{sec: 3}

Under the central technical assumption that phenomenological content corresponds to beliefs (\cref{sec: 2}), we now derive mathematical consequences and empirical predictions. The precise consequences and predictions depend on the nature of the correspondence (\cref{eg: nature}; see Table~\ref{tab: phi-predictions}). In turn, empirical testing of these predictions would help test the central technical assumption, or refine it by informing the nature of the correspondence. From this correspondence, we proceed by first characterising phenomenology at a single moment in time, before turning to its temporal dynamics.

\begin{figure}[t!]
    \centering
    \includegraphics[width= 0.7\textwidth]{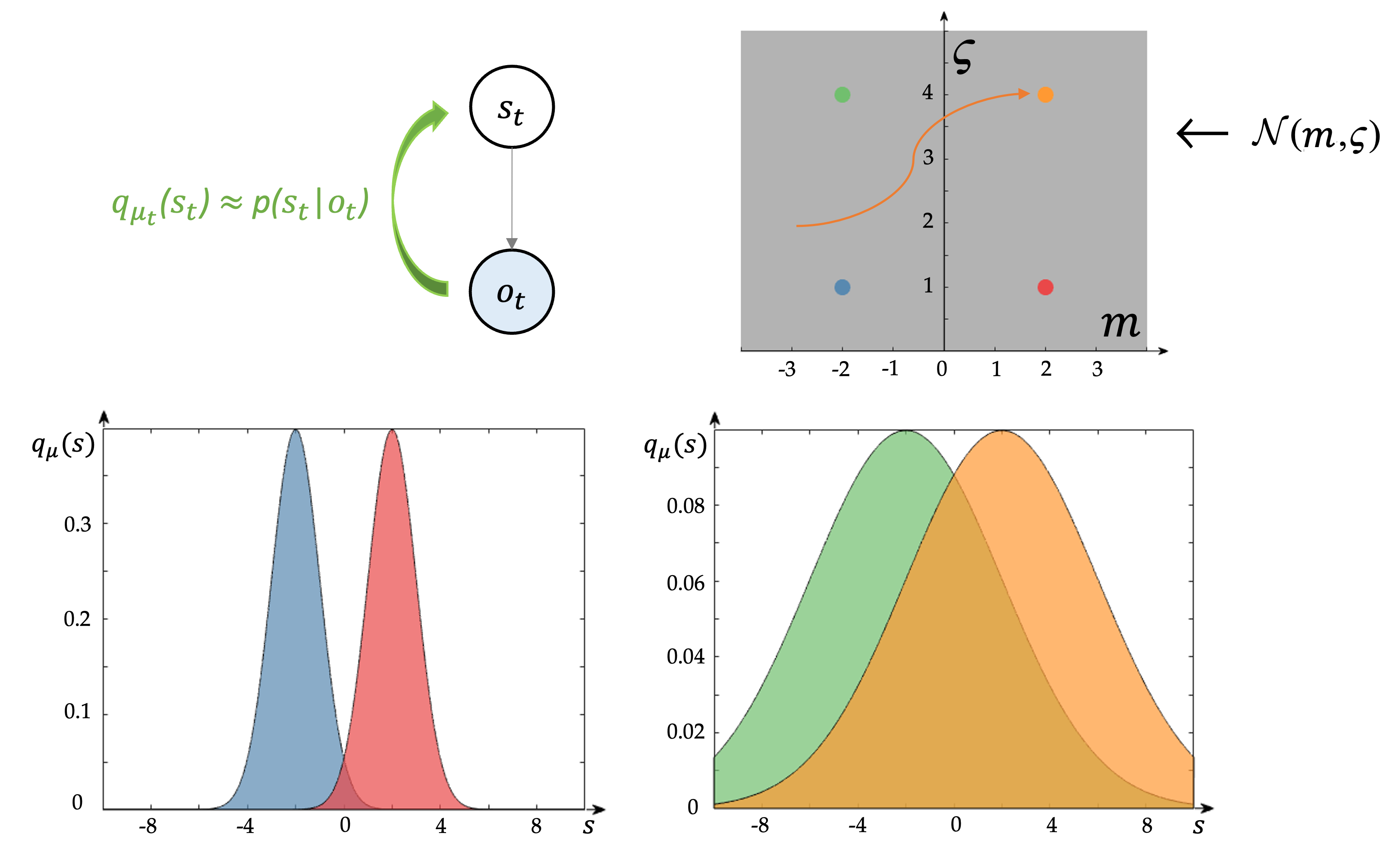}
    \caption{\textbf{Phenomenology and dynamics on the space of beliefs.} This figure showcases phenomenology as a belief about the causes $s$ of our sensory information, e.g. the external temperature. This belief is dynamically updated to approximate a posterior distribution (\emph{top left}). \emph{Bottom:} four subjective beliefs, modelled as Gaussians. Their parameters (mean and standard deviation) are plotted on a two-dimensional half plane (\emph{top right}). The orange arrow illustrates the fact that phenomenology (conceptualised as a belief) changes dynamically; this dynamic can be visualised as a dynamic on the space of parameters.}
    \label{fig: belief dynamics}
\end{figure}

\subsection{A Snapshot of Phenomenology}
\label{sec: snapshot}

What follows is an approach to mathematically describing phenomenology at a single point in time---a snapshot of experience. We start by addressing the question: given two beliefs (whether held by two different individuals, or by the same individual at different times), how can we characterise their difference?

Two distinct notions of difference are relevant here. \emph{Mathematical} differences concern how two beliefs differ in their information content, characterised using information geometry. \emph{Subjective} differences concern how similar or different experiences \emph{feel} to the experiencer themselves---as expressed in a similarity judgement. A key empirical question is whether and when these two notions coincide. We address mathematical characterisation first (\cref{sec: mathematical}), before turning to subjective characterisation and the empirical predictions that follow from hypothesising a relationship between the two (\cref{sec: subjective}).

\subsubsection{Mathematical Characterisation through Information Geometry}
\label{sec: mathematical}

Our goal is to mathematically quantify how beliefs differ in their information content to enable the precise characterisation of phenomenological differences between subjects via \cref{as: core}. As a running phenomenological example, we aim to quantify how similarly two subjects experience the current temperature. To quantify how beliefs differ, we need some measure of discrepancy between them. Any divergence would serve this purpose, noting that all distances are themselves divergences \parencite{ayInformationGeometry2017,barpGeometricMethodsSampling2022}.

The naive approach of computing Euclidean distance between the parameters of beliefs may not be ideal for our purposes, as it fails to capture differences in information content. To illustrate, consider four individuals with Gaussian beliefs about the temperature as in Figure~\ref{fig: belief dynamics}: two believe it is $-2^\circ$C and two believe it is $2^\circ$C, but they differ in their confidence. In parameter space (mean and standard deviation), some belief pairs appear equidistant, yet their beliefs share very different amounts of information---confident beliefs that disagree are more distinct than uncertain beliefs that disagree. We would like a measure of discrepancy that captures this \emph{informational} difference; a distance---satisfying symmetry and the triangle inequality---would be mathematically convenient, though not required.

\paragraph{Information length and Fisher distance.}
Here, we focus on the \emph{Fisher information distance} \parencite{costaFisherInformationDistance2015}, which is natural for several reasons: it measures informational distance (it is the Riemannian distance arising from the Kullback-Leibler divergence, see appendix \ref{app: fisher}), it is invariant under reparameterisation of the belief space, it has deep connections to thermodynamics that we leverage later (\cref{sec: dyn phen}), and it enables the toolbox of Riemannian geometry to be applied. Unlike the KL divergence, which is asymmetric, the Fisher metric is symmetric and defines a proper distance (see Appendix \ref{app: fisher} for the derivation). Other divergences or metrics would also be valid for mathematical characterisation; the Fisher metric is presented here as a natural choice rather than the uniquely correct one.

Intuitively the \emph{information length} of a path through belief space is the accumulated KL divergence along infinitesimal increments of that path (up to a constant transformation, see Appendix \ref{app: fisher}). It quantifies the computational cost of belief updating---the number of natural units of information (nats\footnote{$1$ nat $= \log_2(e)$ bits $\approx 1.44$ bits.}) by which beliefs change along that trajectory. Given a time-differentiable trajectory of beliefs $t \mapsto q_{\mu_t}$ for $t \in [0,1]$, the information length is
\begin{equation}
    \begin{split}
        \ell &= \int_0^1 \sqrt{\dot\mu_t \cdot \left.\nabla_{d\mu}^2 \dkl[q_{\mu_t} \mid q_{\mu_t +d\mu}] \right|_{d\mu = 0} \dot\mu_t}\:dt,
    \end{split}
\end{equation}
where the Hessian matrix in the integrand is the Fisher information metric. The \emph{Fisher information distance} between two beliefs is then defined as the minimal (technically infimal) information length of paths connecting them \parencite{costaFisherInformationDistance2015}.

The Fisher distance admits closed-form expressions for common distributions. For univariate Gaussian beliefs $\mathcal N (m, \varsigma)$ with mean $m$ and standard deviation $\varsigma$, we have \parencite[eq.~9]{costaFisherInformationDistance2015}:
\begin{equation}
\begin{split}
&d\left(\mathcal N\left(m_1, \varsigma_1\right),\mathcal N \left(m_2, \varsigma_2\right)\right) \\
& =\sqrt{2} \ln \left(\frac{\sqrt{\left(\left(m_1-m_2\right)^2+2\left(\varsigma_1-\varsigma_2\right)^2\right)\left(\left(m_1-m_2\right)^2+2\left(\varsigma_1+\varsigma_2\right)^2\right)}+\left(m_1-m_2\right)^2+2\left(\varsigma_1^2+\varsigma_2^2\right)}{4 \varsigma_1 \varsigma_2}\right).
\end{split}
\end{equation}
Returning to our temperature example (Figure~\ref{fig: belief dynamics}), we can now compute the informational differences between the subject's beliefs:
\begin{equation}
\begin{split}
    d[q_{\tikz\draw[blue,fill=blue] (0,0) circle (.4ex);} \mid q_{\tikz\draw[red,fill=red] (0,0) circle (.4ex);}] &=d\left(\mathcal N \left(-2, 1\right),\mathcal N \left(2, 1\right)\right)=\sqrt{2}\log (2\sqrt{6}+5) \approx 3.242 \text{ nats}, \\
    d[q_{\tikz\draw[green,fill=green] (0,0) circle (.4ex);} \mid q_{\tikz\draw[orange,fill=orange] (0,0) circle (.4ex);}] &=d\left(\mathcal N \left(-2, 4\right),\mathcal N \left(2, 4\right)\right) = \sqrt{2}\log (2)  \approx 0.980 \text{ nats}.
\end{split}
\end{equation}
The beliefs of the confident Blue and Red persons are more than three times as different as those of the uncertain Green and Orange persons, even though the means differ by the same amount in both cases. This illustrates a key point: the Fisher distance depends sensitively on precision, not just on the content (mean) of beliefs. Confident beliefs that disagree are more distinct than uncertain beliefs that disagree.

For categorical distributions $q_{\mu}(s)=\operatorname{Cat}(s \mid \mu)$ where $\mu$ is a finite-dimensional vector of non-negative entries that sum to one, the Fisher distance takes the simpler form \parencite[Appendix]{dacostaNeuralDynamicsActive2021}:
\begin{equation}
    d\left(\operatorname{Cat}(s \mid \mu), \operatorname{Cat}(s \mid \mu')\right)= 2\left\|\sqrt{\mu}-\sqrt{\mu'}\right\|.
\end{equation}

\paragraph{Information geometry.}
Looking forward, there is not only a notion of distance available for beliefs but an entire geometry. The Fisher information metric is a Riemannian metric, so one may compute angles, projections, curvature, geodesics, and much more \parencite{amariInformationGeometryIts2016,amariMethodsInformationGeometry2007,ayInformationGeometry2017}. Combined with the additional structure of probability spaces, this yields a rich information-geometric toolbox that may prove fruitful for future work characterising phenomenological differences.

\paragraph{Relationship to phenomenology.}
The space of beliefs $\mathcal{Q}$ comes equipped with a natural geometric structure: the Fisher information metric, with associated distance $d$ and path length $\ell$ as defined above. A central question is whether this structure can illuminate a geometry for phenomenological space $\mathcal{P}$. Under the central technical assumption (\cref{as: core}), the mapping $\varphi \colon \mathcal{Q} \to \mathcal{P}$ provides precisely this bridge---it allows us to \emph{induce} geometric structure on phenomenological space from the well-characterised geometry of belief space. To distinguish the two spaces, we write $d_{\mathcal{Q}}$ and $\ell_{\mathcal{Q}}$ for distances and lengths on belief space, and $d_{\mathcal{P}}$ and $\ell_{\mathcal{P}}$ for their phenomenological counterparts.

How the geometry on $\mathcal{P}$ can be defined, and how it relates to the geometry on $\mathcal{Q}$, depends on the nature of $\varphi$. We distinguish two constructions in Appendix \ref{app: induced geom}. (1) Quotient geometry which is valid for any $\varphi$ (e.g. 1--4 in \cref{eg: nature}). (2) Fisher geometry when $\varphi$ arises as a push-forward (e.g. 1--3 in \cref{eg: nature}). The former is more general but the latter preserves the Fisher metric and its associated Riemannian structure, so it is both mathematically richer and more natural in the context of this work. Both of these geometries need not coincide and they are ultimately a modelling choice. In both cases however, they provide phenomenological distances and path lengths for phenomenology that satisfy the following bounds:
\begin{equation}
\label{eq: dpi bounds}
    d_{\mathcal{P}}(\varphi(q_1), \varphi(q_2)) \leq d_{\mathcal{Q}}(q_1, q_2), \qquad \ell_{\mathcal{P}} \leq \ell_{\mathcal{Q}},
\end{equation}
with equality when beliefs equal phenomenology (1 in \cref{eg: nature}). These bounds tell us that phenomenological distances so-defined cannot exceed the information-theoretic differences in the underlying beliefs. In summary, a mapping $\varphi$ and a choice of geometry on $\mathcal P$ provides a mathematical way to measure differences in experience, and to quantify how differently two subjects experience the current temperature. But how is this mathematical characterisation useful?

\paragraph{Application areas.} In this framework, the geometry of beliefs allows us to precisely characterise phenomenological differences between subjects.\footnote{See \url{https://perceptioncensus.dreamachine.world/} for a large-scale experimental project collecting data on this topic.} For example, a person's phenomenology could be characterised by occupying a characteristic \emph{region} of phenomenological space---under a given set of stimuli. Neurotypicality here could be characterised as belonging to a \emph{region} rather than a single \emph{state}, underlying the fact that there are many ways of being neurotypical and that perceptual diversity is likely a widespread if under-appreciated phenomenon \parencite{sethBigIdeaWe2022}. This framework may also have implications for computational psychiatry where aberrant phenomenology (such as delusions) could be characterised as lying outside the typical region, and mathematical proximity to certain atypical phenomenologies could inform targeted treatments. This is a natural next step for computational psychiatry, which already models psychiatric experiences as aberrant beliefs \parencite{adamsComputationalAnatomyPsychosis2013,adamsComputationalPsychiatryMathematically2015}.
With these mathematical tools in place, we now turn to subjective characterisation: how such differences are experienced and reported.

\subsubsection{Subjective Characterisation and Empirical Predictions}
\label{sec: subjective}

So far we have characterised \emph{mathematical} differences in phenomenology by induction from the mapping $\varphi \colon \mathcal{Q} \to \mathcal{P}$, which gives us a way to ask: given two experiences, how different are they in their information content? A related but distinct question concerns \emph{subjective} differences: given two experiences, how similar or different do they \emph{feel} to the experiencer? Investigating the structure of subjective similarity would provide another lens on possible geometry on $\mathcal{P}$. What follows regarding subjective differences is more speculative but suggests directions for empirical work.

\paragraph{Alternative geometries for subjective similarity.}
Subjective similarity judgments need not obey the axioms of a metric. \textcite{tverskyFeaturesSimilarity1977} noted that such judgments may violate both symmetry (A judged more similar to B than B to A) and the triangle inequality, suggesting that a divergence that is not a distance may be more appropriate for quantifying subjective phenomenological differences. Existing approaches account for these metric violations using quantum geometry \parencite{pothos_quantum_2013,eppingQuantumGeometricFramework2023} or the hypothesis that similarity is computed as an exponentially decaying function of distance \parencite{shepardUniversalLawGeneralization1987}. An interesting future direction would be to use empirical similarity judgments to \emph{infer} the divergence that best describes how the brain quantifies dissimilarities between percepts \parencite{eppingQuantumGeometricFramework2023}, starting with the KL divergence. This complements standard psychophysical approaches such as multidimensional scaling, which infers a low-dimensional embedding where Euclidean distances best match subjective judgments \parencite{torgersonMultidimensionalScaling1952,kruskalMultidimensionalScaling1964}, and maximum likelihood difference scaling (MLDS), which estimates perceptual scales from comparative judgments about which stimulus pairs differ more \parencite{maloneyMaximumLikelihoodDifferenceScaling2003}. Our framework predicts systematic changes in such recovered geometries under manipulations of precision (e.g.\ attention and confidence).

\paragraph{Testable predictions.}
\textit{If} we assume a correlation between mathematical differences in phenomenology (as measured with phenomenological distance $d_{\mathcal P}$) and subjective differences in percepts, \textit{then} we obtain testable predictions. For example, the following predictions necessarily hold under identity and marginal forms for $\varphi$ (\cref{eg: nature}.1-2):
\begin{enumerate}[noitemsep]
    \item If attention modulates the precision of posterior beliefs \parencite{feldmanAttentionUncertaintyFreeEnergy2010,sandved2021towards,parrWorkingMemoryAttention2017}, then unattended stimuli will correspond to less precise beliefs and hence smaller Fisher distances. The prediction is that two stimuli should be judged as \emph{less} distinct when unattended than when attended---testable using similarity judgments under dual-task conditions where attention can be selectively withdrawn \parencite{kawakitaMyRedYour2024,eppingQuantumGeometricFramework2023}.
    \item If the precision of beliefs is reflected in subjective confidence \parencite{parrUncertaintyEpistemicsActive2017,flemingAwarenessInferenceHigherorder2020,geurtsSubjectiveConfidenceReflects2022}, then high-confidence percepts correspond to more precise beliefs and larger Fisher distances. The prediction is that percepts should be judged as \emph{more} distinct when confidence is higher, even for the same stimuli.
\end{enumerate}
Both predictions are amenable to psychophysical experiments, and may be empirically contrasted under alternative forms of $\varphi$ and alternative geometries.

\subsection{Phenomenology over Time}
\label{sec: dyn phen}

Having discussed phenomenology at a single point in time, we now turn to its temporal dynamics. The information length introduced above (\cref{sec: mathematical}) is a natural tool for this purpose: it quantifies `how much' experience changes over time. We examine two applications: subjective cognitive effort and the phenomenology of time.

\subsubsection{Metabolic Cost and Subjective Cognitive Effort}
\label{sec: cost}

\paragraph{Information length and cognitive metabolic cost.}
The information length of a belief trajectory is a geometric measure of how far beliefs move over a finite time window. In physical implementations of inference, finite-time \emph{thermodynamic speed limits} relate the rate of belief change to a minimum degree of thermodynamic irreversibility---typically quantified by \emph{entropy production}\footnote{Often up to an activity/timescale- dependent prefactor.} \parencite{shiraishiSpeedLimitClassical2018,itoStochasticTimeevolutionInformation2020}. In approximately constant-temperature settings (a sensible approximation for brains), greater entropy production is associated with greater energetic dissipation (as heat), so larger (or faster) belief updates can increase the thermodynamic lower bound on energetic dissipation for a fixed task duration. Intuitively, this aligns with the general idea behind Landauer's principle: that informational change can carry an irreducible thermodynamic cost \parencite{landauerIrreversibilityHeatGeneration1961}. Living organisms operate far above these theoretical minima, but one may still expect associations between (i) information length of belief trajectories, (ii) irreversibility of the neural dynamics encoding those beliefs (measured via entropy production), and (iii) brain metabolic expenditure. Consistent with this, information length has been related to entropy production in specific classes of stochastic dynamical systems \parencite{guel-cortezRelationsEntropyRate2023}, and
recent work reports that cognitive load in working memory (controlling for response frequency) and cognitive performance (e.g.\ error rate) correlate with estimated entropy production in neural population dynamics \parencite{lynnBrokenDetailedBalance2021}. 

This motivates the theoretically grounded prediction that---at fixed task duration---greater belief information length should be associated with greater neural entropy production and higher brain metabolic cost, up to an unknown efficiency factor.
Incorporating the association between phenomenological and belief trajectories from \cref{sec: mathematical} yields the predicted associations \cref{eq: diagram metabolic cost}, where solid arrows denote theoretically supported associations in some settings while dashed arrows denote subsequently inferred associations.
\begin{equation}
    \label{eq: diagram metabolic cost}
      \begin{tikzcd}
     &&  \text{Information length } \substack{\\\text{(beliefs)}} \arrow[rd, leftrightarrow, bend left] \arrow[ld, leftrightarrow, bend right] \arrow[dd, leftrightarrow, dashed] & \\
     &\text{Information length } \substack{\\\text{(phenomenology)}}\arrow[dr, leftrightarrow, dashed, bend right] \arrow[rr, leftrightarrow, dashed]&  & \text{Entropy production } \substack{\\\text{(neural populations)}}\\
    &&\text{Cognitive metabolic cost}\arrow[ur, leftrightarrow, bend right]&
        \end{tikzcd}
\end{equation}

\paragraph{Subjective cognitive effort hypothesis.}
\cref{eq: diagram metabolic cost} provides a theoretically predicted bridge from how much phenomenology changes over time (measured with information length) to objective energetic measures; yet a key empirical question is how these quantities relate to \emph{subjective} cognitive effort. We posit the hypothesis that subjective cognitive effort tracks the information length of phenomenological trajectories. This yields a closely related but distinct prediction to existing accounts that operationalise effort via an information length proxy, such as the KL divergence between prior and posterior beliefs \parencite{zenonInformationtheoreticPerspectiveCosts2019,parrCognitiveEffortActive2023}: KL captures the endpoints of a trajectory, whereas information length is trajectory-dependent and accumulates incremental phenomenological change during belief updating. A direct test would measure subjective effort ratings and infer belief and phenomenological trajectories in the same task, and compare how well (i) information length, (ii) prior--posterior KL, and (iii) objective energetic measurements (metabolic expenditure and/or neural irreversibility) predict reported cognitive effort.
\begin{equation}
    \begin{tikzcd}
    \text{Subjective cognitive effort}\arrow[rrrrrr, leftrightarrow, dashed]&&&&&&\text{Information length } \substack{\\\text{(phenomenology)}}
    \end{tikzcd}
\end{equation}

\subsubsection{Phenomenology of Time}
\label{sec: time}

We now turn to another example application: the phenomenology of time perception, and specifically, the experience of temporal duration. The hypothesis here is that the information length of phenomenology may be apt for quantifying the subjective experience of duration.

\paragraph{Existing approaches.}
Time perception has traditionally been explained by appealing to inner `clocks' that track objective time \parencite{churchPropertiesInternalClock1984,meckNeuropharmacologyTimingTime1996,wittmannInnerExperienceTime2009,eaglemanTimeBrainHow2005}.
More recently, an alternative proposal has emerged, which argues that subjective duration can be accounted for by accumulated salient change in perceptual processing \parencite{roseboomActivityPerceptualClassification2019}.
Roseboom and colleagues exposed a pre-trained image classification network (AlexNet \parencite{krizhevskyImageNetClassificationDeep2012}) to video snippets, and modelled subjective time by accumulating the number of times dynamic `salience thresholds' were crossed at various successive stages in the network \parencite{roseboomActivityPerceptualClassification2019}.
In this model, salience is measured by the Euclidean distance between successive activation patterns within a given layer of the network, and a unit of subjective time is accumulated whenever this salience metric exceeds the arbitrary threshold at any given layer \parencite[p2]{roseboomActivityPerceptualClassification2019}.
Furthermore, attention is seen as modulating this salience threshold: low attention means a high salience threshold: when we are not paying attention to something, we are less likely to notice it changing, but large changes will still be noticed, and vice versa. In particular, low attention entails shorter subjective durations, and vice versa. This model was able to accurately predict human duration judgements of the same videos, including characteristic biases (over-estimating short durations and underestimating long durations). Notably, accurate predictions were still possible when model activity was substituted by corresponding perceptual brain activity recorded in fMRI \parencite{shermanTrialbytrialPredictionsSubjective2022}, suggesting that the model is picking out relevant features of neural activity, and therefore constitutes a form of computational phenomenology.

Here, we propose an alternative account of these findings using information length.

\paragraph{Salience as information gain.}
In predictive processing, one notion of (epistemic) \emph{salience} of an observation $o$ is the \emph{information gain} it affords about latent causes $s$ \parencite{mirzaSceneConstructionVisual2016}. Mathematically, this is the KL divergence between the posterior belief following an observation (say at time $t$) and the belief prior to the observation (say at time $t-1$):
\begin{equation*}
    \overbrace{ \underbrace{\dkl[q_{\mu_t}(s) \mid q_{\mu_{t-1}}(s)]}_{\text{Information gain}}}^{\text{Salience}}
\end{equation*}
In other words, the degree to which an observation is salient is the extent to which the associated beliefs move following this observation. Counting salient observations thus corresponds to measuring the rate at which beliefs travel through belief space. Since the beliefs that are modelled as such in the literature are usually consciously experienced, it follows that this also measures the extent to which phenomenology changes over time.

\paragraph{Subjective time as information length.} Consistent with this, we propose two hypotheses: that subjective time associated with experiencing a sequence of stimuli corresponds to the information length of beliefs---respectively of phenomenology---as successive stimuli impinge. If stimuli are salient, beliefs and phenomenology change further, and subjective time will be large---and vice versa. This proposal complements the Roseboom approach while offering three advantages. First, it requires no arbitrary salience thresholds: salience is naturally accumulated with information length. Second, the role of attention is intrinsic rather than requiring a separate threshold-modulating mechanism. Third, it furnishes testable hypotheses in terms of subjective confidence.

\paragraph{Role of attention.}
If attention modulates the precision of posterior beliefs \parencite{feldmanAttentionUncertaintyFreeEnergy2010,sandved2021towards,parrWorkingMemoryAttention2017}, high attention yields more precise beliefs, which incur larger information lengths when they change (\cref{sec: mathematical}). Consider again Figure~\ref{fig: belief dynamics}: an attending subject (Blue distribution) and a non-attending subject (Green distribution) both experience the temperature rising from $-2^\circ$C to $2^\circ$C. The attending subject's beliefs shift from Blue to Red, accumulating a larger information length than the non-attending subject's shift from Green to Orange. Thus, the attending subject experiences more subjective time---consistent with the common observation that attended events feel longer. Note that whether precise belief changes necessarily accumulate larger phenomenological lengths depends on the nature of $\varphi$; this is the case under identity and marginal forms (\cref{eg: nature}.1-2), but not necessarily under (\cref{eg: nature}.3-4). Hypothesising the nature of $\varphi$ therefore provides complementary and potentially contrastive predictions for disambiguating the role of information length of phenomenology and its relationship to subjective time.

\paragraph{Predictions from subjective confidence.} If the precision of beliefs is reflected in subjective confidence \parencite{parrUncertaintyEpistemicsActive2017,flemingAwarenessInferenceHigherorder2020,geurtsSubjectiveConfidenceReflects2022}, then higher confidence corresponds to more precise beliefs and larger information distances. If the information length of beliefs corresponds to subjective time, the prediction is that subjective time should feel longer when subjective confidence is higher, and vice-versa. If on the other hand, the information length of phenomenology corresponds to subjective time, the prediction is that under identity and marginal forms of $\varphi$ (\cref{eg: nature}.1-2) subjective time should feel longer when subjective confidence is higher. This is because the relationship between precision and information length carries over under such belief-phenomenology correspondence. Under more generic correspondences (\cref{eg: nature}.3-4) the relationship between belief precision and information length needs to be established on a case-by-case basis.

\paragraph{Future empirical directions.}
While the proposal advanced here lacks the detail and engagement with empirical data of the Roseboom et al studies, it offers a complementary perspective and new empirical predictions. It would be interesting to compare the two approaches using the same data. Furthermore, hierarchical generative models \parencite{fristonHierarchicalModelsBrain2008} may help account for different granularities of time perception; where we seem to experience duration differently over different time scales \parencite{singhalTimeTimeAgain2021}. Keeping track of both short time-spans and long time-spans simultaneously could possibly be modelled as the information length accrued at different levels of the model's hierarchy, extending related work in time perception \parencite{singhalTimeTimeAgain2021,shermanTrialbytrialPredictionsSubjective2022,fountasPredictiveProcessingModel2022,roseboomActivityPerceptualClassification2019}.

\section{Beliefs and Neural Dynamics}
\label{sec: 4}

Having developed the connection between beliefs and first person experiential reports under the central technical assumption (\cref{as: core}), we review an emerging connection between beliefs and neural dynamics (\cref{fig: rosetta stone}, top-right) completing a proposed generative passage between neural recordings and experiential reports. Our aim is not a comprehensive review, but a proof of concept for completing a generative passage between experiential reports and neural recordings. We focus on the connection between neural and belief dynamics under partially observed Markov decision process generative models (POMDPs)---noting that other connections are possible under other types of (e.g.\ continuous state space) generative models \parencite{fristonTheoryCorticalResponses2005,fristonPredictiveCodingFreeenergy2009,fristonGraphicalBrainBelief2017}. The connections we review describe neural processes as engaging in variational Bayesian inference about the causes of their sensory input by optimising an evidence lower bound \parencite{knillBayesianBrainRole2004a,fristonFreeenergyBrain2007,fristonFreeenergyPrincipleUnified2010,isomuraCanonicalNeuralNetworks2022,isomuraExperimentalValidationFreeenergy2023} (see also \cref{app: pp}).

\begin{figure}[t!]
    \centering
    \includegraphics[width= 0.4\textwidth]{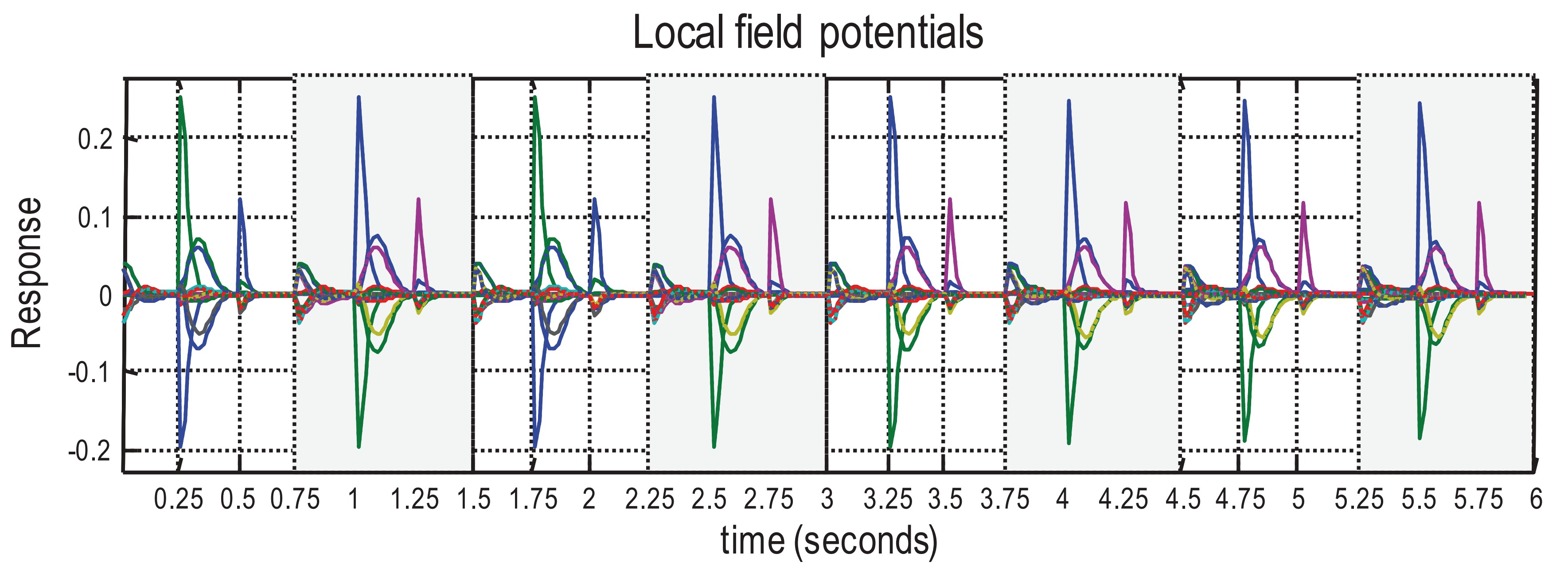}
    \caption{\textbf{Simulated neural population dynamics.} This figure shows simulated local field potentials under active inference accounts of predictive processing. These are simulated from belief dynamics, as an organism samples a sequence of stimuli. For more details on these simulated dynamics, see \parencite{fristonActiveInferenceProcess2017,dacostaNeuralDynamicsActive2021,isomuraCanonicalNeuralNetworks2022}.}
    \label{fig: LFP}
\end{figure}

\subsection{From beliefs to neural dynamics}
\label{sec: belief to neural}

First, we go from belief dynamics to neural recordings: how does the process of updating one's beliefs via variational inference correspond to neural dynamics? Here we review some active inference accounts of predictive processing that propose hypothetical neural population dynamics from variational inference equations in POMDPs \parencite{fristonActiveInferenceProcess2017,dacostaActiveInferenceDiscrete2020,dacostaNeuralDynamicsActive2021}.

\paragraph{Belief dynamics.} We consider the simplest example where an organism is described as representing some of its environment in terms of a finite number of possible states (e.g., locations in space encoded by place cells) using a POMDP \parencite{astromOptimalControlMarkov1965a,dacostaActiveInferenceDiscrete2020}. When this is the case, one simple hypothesis for its belief dynamics about the current state are the following equations which unfold in peristimulus time \parencite{dacostaActiveInferenceDiscrete2020}
\begin{equation}
\label{eq: first passage}
        \dot{\mu}=-\nabla_{\sigma(\mu)} \mathrm{F}[q_\mu], \qquad q_\mu(s)=\operatorname{Cat}(s\mid \sigma(\mu)).
\end{equation}
In this equation, $\operatorname{F}$ is the variational free energy functional (i.e., negative evidence lower bound \parencite{bleiVariationalInferenceReview2017,bealVariationalAlgorithmsApproximate2003}), $\sigma$ is a softmax function and $ q_\mu$ represents the agent's beliefs about external states. This is a categorical distribution parameterised by $\sigma(\mu)$. Explicitly, $\sigma(\mu)$ is a vector whose \textit{i}-th component is the agent's belief (expressed as a probability) that it is in the \textit{i}-th state. The softmax function is the natural choice to map from parameters to beliefs as the former turns out to have a logarithmic form \parencite[eq. 8]{dacostaActiveInferenceDiscrete2020} and the components of the latter must sum to one.

\paragraph{Neural predictions.} Neurons convert post-synaptic voltage potentials to firing rates just as these dynamics convert a vector of real numbers $\mu$, to a vector whose components are bounded between zero and one $\sigma(\mu)$. Thus it is natural to map $\mu$ as the voltage potential of neuronal populations, and $\sigma(\mu)$ as their firing rates (since these are upper bounded due to neuronal refractory periods). This allows one to simulate a variety of neural responses including local field potentials (\cref{fig: LFP}). We now point towards evidence for this way of thinking. 

\paragraph{Face validity.} The idea that state estimation can be expressed in terms of firing rates is well-established when the state-space constitutes an internal representation of space. This is the raison d'\^etre for the study of place cells \parencite{stachenfeldHippocampusPredictiveMap2017}, grid cells \parencite{haftingMicrostructureSpatialMap2005} and head-direction cells \parencite{chenHeaddirectionCellsRat1994,taubeHeaddirectionCellsRecorded1990}, where the states inferred are (under some perspectives) physical locations in space \parencite{rajuSpaceLatentSequence2024}. Primary afferent neurons in cats have also been shown to encode kinematic states of the hind limb \parencite{steinCodingPositionSimultaneously2004,wagenaarStatespaceDecodingPrimary2011,weberDecodingSensoryFeedback2006}. Most notably, the seminal work of Hubel and Wiesel \parencite{hubelReceptiveFieldsSingle1959} showed the existence of neurons encoding orientation of visual stimuli. In short, the very existence of receptive fields in neuroscience suggests a carving of the world into discrete states under an implicit discrete-state generative model. While many of these studies focus on single neuron recordings, the arguments presented apply equally to populations comprising multiple neurons.

\paragraph{Theoretical and Empirical Evidence.} There are complementary theoretical and empirical research strands supporting the correspondence between state-estimation and neural dynamics reviewed here. This correspondence holds mathematically in a large class of biological neural network models, comprising rate coding models, known as `canonical neural networks' \parencite{isomuraBayesianFilteringMultiple2019}. More generally, it is consistent with mean-field models of neural population dynamics \parencite{moranNeuralMassesFields2013,brunelFiringRateNoisy2003} where the softmax function plays the same role of translating average potentials to firing rates.
In addition, information-geometric arguments similar to \cref{sec: cost} suggest that beliefs dynamics in \eqref{eq: first passage} are computationally and metabolically efficient, predicting that the neural processes implementing them are also efficient, consistently from what we would expect from real neurons, where efficiency has been naturally selected for throughout evolution \parencite{senguptaInformationEfficiencyNervous2013}.
Finally, the reviewed correspondence allows one to synthesise a wide range of biologically plausible electrophysiological responses, including local field potentials, repetition suppression, mismatch negativity, violation responses, place-cell activity, phase precession, theta sequences, theta-gamma coupling, evidence accumulation, race-to-bound dynamics and transfer of dopamine responses \parencite{fristonActiveInferenceProcess2017,schwartenbeckDopaminergicMidbrainEncodes2015}. These predicted responses have been validated empirically with in-vitro neural networks that self-organised to perform discrete-state inference \parencite{isomuraExperimentalValidationFreeenergy2023}.

\subsection{From neural recordings to beliefs}

Conversely, going from neural recordings to belief updating usually entails \emph{reverse-engineering} the generative model embodied by the organism we are recording from in addition to its belief dynamics.

\paragraph{Canonical neural networks as backbone.} As mentioned in \cref{sec: belief to neural}, a large class of biological neural network models known as canonical neural networks can be described as performing variational inference on POMDPs via \eqref{eq: first passage} \parencite{isomuraCanonicalNeuralNetworks2022}. Additionally, the parameters of canonical neural network models have been shown to be in one-to-one correspondence with the priors of POMDPs. For instance, firing thresholds correspond to hidden state and decision priors. In other words, different parameterisations of network dynamics correspond to belief updating under the same POMDP with different prior beliefs \parencite{isomuraCanonicalNeuralNetworks2022}. These foundations can be helpful for reverse engineering generative models and belief updates from neural recordings \parencite{isomuraExperimentalValidationFreeenergy2023}.

\paragraph{From real recordings.} This mathematical backbone was applied to \textit{in vitro} neural network recordings from rat cortical neurons. \textcite{isomuraExperimentalValidationFreeenergy2023} developed a technique for reverse-engineering the parameters of POMDPs (including prior beliefs) from neural recordings following sensory stimuli (see also \parencite{isomuraVitroNeuralNetworks2018,isomuraCulturedCorticalNeurons2015}). They showed that the variational inference equations on POMDPs implemented by canonical neural networks accurately predict future in-vitro neural responses and the trajectory of synaptic strengths (i.e., learning). Furthermore, they showed that the change in baseline excitability of in vitro networks is consistent with the change in prior beliefs about external states, validating that priors over hidden states are encoded by firing thresholds in this setting. This study reverse-engineering belief dynamics from neural recordings was recently extended to \textit{in vivo} neural networks, from large-scale calcium imaging data of zebrafish, lending additional predictive validity to this setting \parencite{isomuraPredictingIndividualLearning2025}.

These findings suggest that several types of biological neural networks perform variational Bayesian belief updating under a POMDP generative model when the external causes of sensory input are discrete.\footnote{It begs the question whether the same networks of neurons can also self-organise to embody continuous state generative models, when the external states are continuous.} Altogether, this approach shows how it is possible to reverse engineer generative models and the accompanying belief dynamics from neural activity alone.

\section{Discussion}

\paragraph{A method for computational phenomenology.} Core to this work is the methodological assumption that phenomenological content is a function of an organism's beliefs, considered as probability distributions (Assumption \ref{as: core}). This assumption enables the application of predictive processing to phenomenology. While this assumption is plausible, it remains a matter of debate. We have pursued a conditional approach: \emph{If} the assumption holds, \emph{then} certain predictions follow. Following a broadly Lakatosian perspective \parencite{lakatosMethodologyScientificResearch1978}, this method may be considered valuable over time (and credence in the core methodological assumption increased) if these hypotheses turn out to be testable and that testing leads to explanatory insight and predictive ability. If not, the method will become less valuable, and credence in the core methodological assumption lessened. We hope this method is productive in this sense, not degenerate.

\paragraph{Future empirical directions.} Future empirical work should test the specific experimental predictions raised in this paper for (1) subjective similarity judgements (\cref{sec: subjective}), (2) cognitive metabolic cost and subjectively experienced cognitive effort (\cref{sec: cost}), and (3) the experience of temporal duration (\cref{sec: time}), comparing with existing studies, e.g. \parencite{roseboomActivityPerceptualClassification2019}. These experiments will help elucidate the broader connection between beliefs and phenomenology, which beliefs are phenomenological, and the validity of the central technical assumption. To strengthen the generative passage between phenomenology and neural dynamics, future work should also improve the strength and scope of the connection between beliefs and neural dynamics \parencite{kaganVitroNeuronsLearn2022,isomuraExperimentalValidationFreeenergy2023,isomuraVitroNeuralNetworks2018}. Please see \parencite{isomuraCanonicalNeuralNetworks2022,isomuraExperimentalValidationFreeenergy2023,isomuraPredictingIndividualLearning2025} and \parencite[Discussion]{dacostaNeuralDynamicsActive2021} for more details on this ongoing programme.

\paragraph{From bridging principles to theories of consciousness.} The generative passages developed in this work are very much aligned with a `real problem' approach to consciousness, in which---rather than proposing necessary and/or sufficient conditions for consciousness---the idea is to build explanatory bridges between properties of consciousness and properties of mechanism \parencite{sethBeingYouStory2021,hohwyPredictiveProcessingSystematic2020,ramsteadGenerativeModelsGenerative2022,seth_hard_nodate,ramstead_naturalizing_2015,roy_beyond_1999}.
This lays predictive processing as a theory of use for consciousness research rather than a theory of consciousness \textit{as such} \parencite{hohwyPredictiveProcessingSystematic2020}. Other perspectives are possible, where one seeks to identify further, necessary or sufficient conditions for a belief to be part of conscious content \parencite{dacostaBayesianMechanicsMetacognitive2024,ramsteadInnerScreenModel2023,WHYTE20264,laukkonenBeautifulLoopActive2024}. In doing so, it is possible that a core set of theoretical commitments will emerge, and that this set will constitute a predictive processing theory of consciousness \emph{per se}. Whichever way things play out, there is great promise that the mathematical and conceptual tools provided by predictive processing will help expose the neural basis of many different kinds of subjective experience.

\section{Conclusion}

Neurophenomenology seeks to build generative passages between first-person phenomenological descriptions and third-person neuroscientific and behavioural measurements. We have approached this challenge using predictive processing, adopting the central technical assumption that phenomenological content is a function of beliefs. This provides a Rosetta Stone hypothesis where beliefs serve as a hub connecting phenomenology, behaviour, and neural dynamics. Taking a conditional approach---if the assumption holds, then these certain consequences follow---we derived testable predictions for subjective similarity judgements, cognitive metabolic cost, subjective cognitive effort, and time perception. Future experimental work testing these predictions will help elucidate the validity of this central assumption connecting phenomenology with beliefs and advance the computational neurophenomenology programme.

\subsubsection*{Acknowledgements}

This work was supported by a workshop at the Lorentz Centre and a travel stipend by Mind and Life Europe.

\subsubsection*{Funding information}

AKS is supported by the European Research Council (Advanced Investigator Grant ERC-AdG-101019524).

\appendix

\section{Bayesian Mechanics Foundations of Predictive Processing}
\label{app: pp}

Here we briefly review Bayesian mechanics, a branch of physics which suggests that it is not surprising that we can describe a variety of organisms as encoding beliefs about their external states and optimising those beliefs via variational inference---lending a complementary, theoretically grounded foundation for predictive processing.

\begin{figure}[h!]
    \centering
    \includegraphics[width=\textwidth]{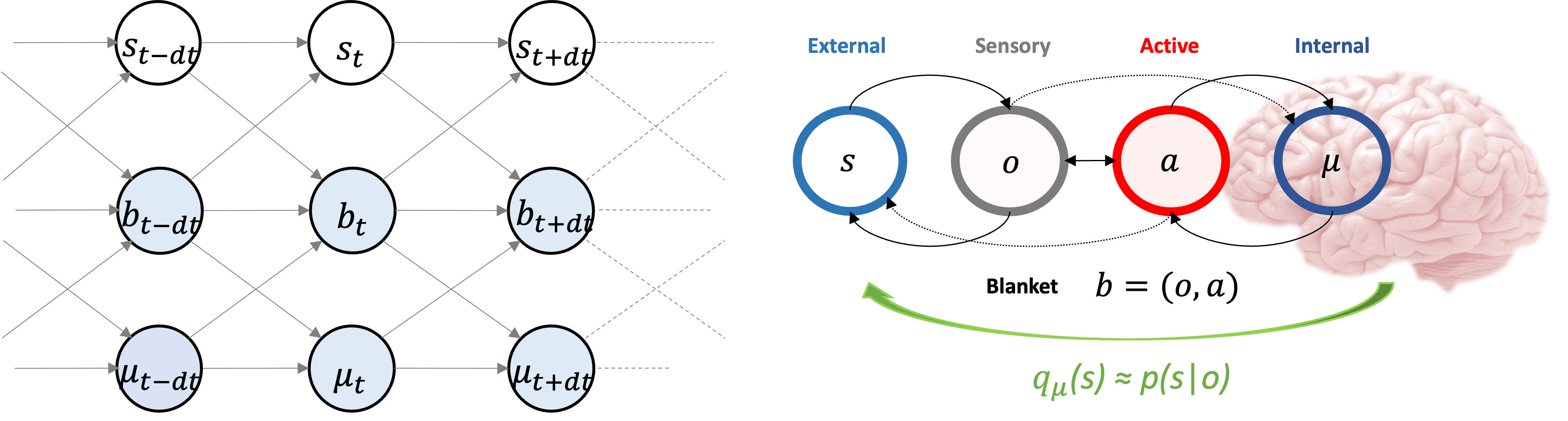}
    \caption{\textbf{Bayesian mechanics.} This figure shows the separation between the dynamically evolving external $s$ and internal $\mu$ states, whereby all interactions are mediated by the boundary or blanket states $b$. \textit{Left}: We see the dynamics evolving over time in a causal network where external variables are in white, while variables that belong to the organism are in blue. \textit{Right}: The Markov blanket is decomposed into sensory (i.e., observations) and active states, operationally defined as those which are not influenced by internal and external states, respectively. The green arrow illustrates that internal states $\mu$ can often be described as encoding beliefs (i.e., probability distributions) $q_\mu(s)$ about external states of the world, which approximate true posterior beliefs given sensory states $o$, and which are updated consistently with variational inference in predictive processing, statistics and machine learning.}
    \label{fig: Markov blanket}
\end{figure}

\subsection{At a high level}

Bayesian mechanics describes the dynamics of entities---defined by possessing a boundary that persists over some interval of time---as inferential processes.\footnote{Note that we do not make any ontological claims in this paper about organisms actually implementing a process of inference; rather, the claim is that their dynamics can be described as a process of inference.}
The common starting point for Bayesian mechanics is a description of the system at hand---comprising the entity and its environment---as a random dynamical system---the conclusion is a description in terms of the internal states of the entity as performing (approximate) Bayesian inference.
\begin{equation*}
  \begin{tikzcd}
    \text{Random dynamical system} \arrow[rrrr, rightarrow, "\text{mathematical theory}"] &  & & &\text{Description as inference}.
    \end{tikzcd}
\end{equation*}
The descriptions as inference usually take the form of a (stochastic) gradient descent on a free energy functional (a.k.a. evidence lower bound), consistent with variational Bayesian inference in statistics, machine learning and theoretical neuroscience \parencite{bleiVariationalInferenceReview2017,bishopPatternRecognitionMachine2006}. These results hold under mild regularity conditions on the nature of certain classes of commonly encountered families of random dynamical systems, e.g., stationary processes \parencite{dacostaBayesianMechanicsStationary2021}, diffusions \parencite{fristonFreeEnergyPrinciple2023a,fristonStochasticChaosMarkov2021}, and Markov chains \parencite{parrMessagePassingMetabolism2021}. The suggestion here is
that belief updating is an emergent property of a wide variety of physical entities in virtue of interacting with their environment via a boundary.

\subsection{In more detail}

An entity---such as a brain or a human---exists over some time interval in virtue of being distinguishable from its surrounding environment during this time \parencite{friston2012free}. This distinguishability entails the existence of a set of states that constitute the entity's boundary which separates and couples it to everything else. A system containing the entity can thus be partitioned into three sets of states: the external states $s$ that belong to the environment, the internal states $\mu$ that belong to the entity, and the blanket states $b$ that constitute the boundary.  Mathematically, the boundary is a Markov blanket between internal and external paths (see Figure \ref{fig: Markov blanket} left). By this we mean that any influence from the external states to the internal states (and vice versa) must occur via blanket states.

The blanket states themselves can be partitioned in terms of the influence they exert on the inside and outside of the entity. The boundary is composed of sensory states $o $ and active states $a $ (which may or may not be empty \parencite{fristonPathIntegralsParticular2023}), where the active states can influence the environment, but not vice versa, and the sensory states can influence the internal states, but not vice versa (see Figure \ref{fig: Markov blanket} right). In this framework, an inert entity such as a rock is simply one that has no active states (but a radioactive rock has active states).

This partition of the system into blanket, internal and external states is known as the `particular partition' (as entities are referred to as `particles' in Bayesian mechanics) \parencite{fristonFreeEnergyPrinciple2019a,fristonFreeEnergyPrinciple2023a}. A particular partition enables, under some conditions, to obtain a mathematically equivalent description of the dynamics of the entity as performing inference over the external states $s$ given its sensory states. Specifically, we mean that internal states parameterise beliefs about (i.e., probability distributions over) external states \parencite{dacostaBayesianMechanicsStationary2021}, so that they become estimators of external states \parencite{ramstead2024approach}.\footnote{We emphasise that the term `belief' is used here in a statistical sense, which not necessarily equivalent to the sense of the term as used in philosophy, to denote a propositional attitude with truth conditions \parencite{smith2022active}.}
For example, given a fixed sensory state, there is a mapping from internal states to approximate posterior beliefs about external states, such that the belief corresponding to the most likely internal state approximates the true posterior, given the sensory state \parencite{fristonFreeEnergyPrinciple2023a, ramsteadBayesianMechanicsPhysics2022}:
\begin{equation}
    \mu \mapsto q_\mu(s) \approx p(s \mid o).
\end{equation}
Here $p$ can be the stationary solution to the density dynamics, i.e., the non-equilibrium steady state of the process describing the system \parencite{dacostaBayesianMechanicsStationary2021,fristonFreeEnergyPrinciple2023a,ramsteadBayesianMechanicsPhysics2022}, or the (typically non-stationary) distribution of the system over paths \parencite{ramsteadBayesianMechanicsPhysics2022,fristonPathIntegralsParticular2023}, in which case the belief $q_\mu$ is also taken over external paths. The nature of beliefs can vary from entity to entity depending on its dynamical properties, from simple to complex \parencite{fristonPathIntegralsParticular2023} and from structured to unstructured \parencite{dacostaBayesianMechanicsMetacognitive2024,sandved-smithMetacognitiveParticlesMental2024}. Importantly, inference depends on where we draw the entity's boundary: every organism, even a cell, has its own boundary, and complex organisms like ourselves are thought to be formed of nested boundaries at multiple spatial scales \parencite{kirchhoffMarkovBlanketsLife2018}. This means that the brain can entertain beliefs about the body and the body's environment, and brain regions can entertain beliefs about other brain regions \parencite{dacostaBayesianMechanicsMetacognitive2024,sandved-smithMetacognitiveParticlesMental2024}---a perspective that accommodates interoceptive inference \parencite{sethInteroceptiveInferenceEmotion2013,sethActiveInteroceptiveInference2016}.

In this setup, it follows in a variety of cases, that the internal and active states evolve based on incoming sensory data by minimising variational free energy \parencite{dacostaBayesianMechanicsStationary2021,fristonFreeEnergyPrinciple2023a,fristonStochasticChaosMarkov2021}

\begin{equation}
a, \mu \searrow \operatorname F\left[q_\mu,o\right]:=\underbrace{\dkl\left[q_\mu(s) \mid p(s \mid o)\right]}_{\text{Bayesian brain}}\underbrace{-\log p(b, \mu)}_{\text{Self-evidencing}}.
\end{equation}

The first term in the variational free energy is the discrepancy between the beliefs that the entity has about the external states and the posterior belief---as measured with the Kullback-Leibler (KL) divergence \parencite{kullbackInformationSufficiency1951}. Minimising this divergence ensures that the beliefs of the entity about its environment are continuously updated in light of the available sensory data. The second term---$p(b, \mu)$---is the evidence for the states of the organism if we interpret $p(s, b, \mu)$ as a generative model of how external states influence the states of the entity; i.e., a generative model for how the environment affects the organism. In other words, internal and active dynamics maximise the evidence for the entity---a description known in philosophy as self-evidencing \parencite{hohwySelfEvidencingBrain2016}
\begin{align*}
    \underbrace{p(s \mid b, \mu)}_{\text{Posterior}}=\frac{\overbrace{p(b, \mu \mid s)p(s)}^{\text{Generative model}}}{\underbrace{p(b, \mu)}_{\text{Evidence}}}.
\end{align*}

In conclusion, a variety of persistent entities can be described as encoding beliefs about their external states that evolve by minimising free energy to make sense of incoming sensory data. Note that we have not described here the precise conditions under which this family of results holds. Although some work focused on deriving precise conditions for simple classes of stationary processes \parencite{dacostaBayesianMechanicsStationary2021}, and deriving these results for specific systems \parencite{fristonStochasticChaosMarkov2021}, much more remains to be done to precisely derive these conditions in more complex system classes \parencite{fristonFreeEnergyPrinciple2023a,fristonPathIntegralsParticular2023,parrMessagePassingMetabolism2021,sandved-smithMetacognitiveParticlesMental2024,dacostaBayesianMechanicsMetacognitive2024}.

\subsection{Active inference}

Bayesian mechanics underwrites a framework to model and simulate the internal and active dynamics of organisms known as `active inference' \parencite{dacostaActiveInferenceDiscrete2020,parrActiveInferenceFree2022,smith2022active,bogaczTutorialFreeenergyFramework2017,buckleyFreeEnergyPrinciple2017,vanoostrumConciseMathematicalDescription2024,fristonFreeEnergyPrinciple2023a}. Active inference is the converse of Bayesian mechanics: one specifies a generative model of how the external world causes the sensory states of an organism, then simulates the ensuing cognitive and behavioural processes (perception, learning, action, etc.) by minimising free energy \parencite{fristonFreeenergyPrincipleUnified2010}. In this sense, the generative model does all of the heavy lifting: what differentiates different organisms is their generative model and the observations used to invert it.

\section{Derivation of the Fisher Information Metric}
\label{app: fisher}

The Kullback-Leibler (KL) divergence is a privileged measure of discrepancy between probability distributions. However, the KL divergence is not a distance because it is asymmetric:
\begin{equation}
    \dkl[q_{\tikz\draw[red,fill=red] (0,0) circle (.4ex);} \mid q_{\tikz\draw[blue,fill=blue] (0,0) circle (.4ex);}] \neq \dkl[q_{\tikz\draw[blue,fill=blue] (0,0) circle (.4ex);} \mid q_{\tikz\draw[red,fill=red] (0,0) circle (.4ex);}].
\end{equation}
However, when the two distributions are infinitesimally close, the KL divergence becomes symmetric. To see why, let $\mu$ denote the parameters of the probability distributions (e.g., for a Gaussian, $\mu = (m, \varsigma)$ comprises the mean and standard deviation). Consider a second-order Taylor expansion of the KL divergence around its first argument, viewed as a function of a small change in parameters $d\mu$:
\begin{equation}
    \label{eq: taylor}
    \dkl[q_{\mu} \mid q_{\mu +d\mu}] = \underbrace{\left. \dkl[q_\mu \mid q_{\mu +d\mu} ]\right|_{d\mu = 0}}_{=0} +  \underbrace{\left.\nabla_{d\mu} \dkl[q_\mu \mid q_{\mu +d\mu} ] \right|_{d\mu = 0} d\mu}_{=0}+\frac 1 2 d\mu \cdot \underbrace{\left(\left.\nabla_{d\mu}^2 \dkl[q_\mu \mid q_{\mu +d\mu}] \right|_{d\mu = 0} \right)}_{\text{Fisher information metric}}d\mu+ o\left(\|d \mu\|^2\right).
\end{equation}
The leading term vanishes because the KL divergence between identical distributions is zero. The second term also vanishes because the KL divergence is minimised when its arguments are equal. This leaves the third term, which is generally non-zero, symmetric in $d\mu$, and quadratic in the infinitesimal parameter difference. The matrix appearing in this term is the \emph{Fisher information metric}. This defines a Riemannian metric on the space of beliefs.

Intuitively, this Riemannian metric defines a distance that is valid locally (for distributions that are infinitessimally close) by:\footnote{The square root appears because, for infinitesimally close distributions in a smooth family, KL is equivalent to second-order a squared Riemannian distance: \eqref{eq: taylor}.}
\begin{equation}
d(q_{\mu}, q_{\mu +d\mu})= \sqrt{2 \dkl[q_{\mu} \mid q_{\mu +d\mu}]}=\sqrt{d\mu \cdot \underbrace{\left(\left.\nabla_{d\mu}^2 \dkl[q_\mu \mid q_{\mu +d\mu}] \right|_{d\mu = 0} \right)}_{\text{Fisher information metric}}d\mu}.
\end{equation}
The extension from this local definition to a global distance proceeds via path integration: infinitesimal increments of distance can be accumulated over arbitrarily long trajectories, yielding the information length of a path. Given a trajectory of beliefs $t \mapsto q_{\mu_t}$ for $t \in [0,1]$, the information length is
\begin{equation}
    \begin{split}
        \ell &= \int_0^1 d(q_{\mu_t}, q_{\mu_t +d\mu_t})\\
        &=\int_0^1 \sqrt{d\mu_t \cdot \left.\nabla_{d\mu}^2 \dkl[q_{\mu_t} \mid q_{\mu_t +d\mu}] \right|_{d\mu = 0} d\mu_t}\\
        &=\int_0^1 \sqrt{\dot\mu_t \cdot \left.\nabla_{d\mu}^2 \dkl[q_{\mu_t} \mid q_{\mu_t +d\mu}] \right|_{d\mu = 0} \dot\mu_t}\:dt,
    \end{split}
\end{equation}
The latter integral is defined only when the trajectory $\mu_t$ on the parameter space (i.e. statistical manifold) is time-differentiable. The \textit{Fisher information distance} between two beliefs is then the infimal information length of paths connecting them.

\section{Induced Geometry}
\label{app: induced geom}

The central technical assumption (\cref{as: core}) posits a mapping $\varphi \colon \mathcal{Q} \to \mathcal{P}$ from beliefs to phenomenology. This appendix addresses how geometric structure on belief space $\mathcal{Q}$ transfers to phenomenological space $\mathcal{P}$ under this mapping. Two constructions are available, differing in generality and predictive content.

\subsection{Quotient Geometry}
\label{app: induced quotient}

For any surjective map $\varphi \colon \mathcal{Q} \to \mathcal{P}$, one can define a distance between phenomenological states defined as the minimal (i.e. infimal) distance between all belief pairs that give rise to the phenomenological states in question
\begin{equation}
\label{eq: quotient metric}
    d_{\mathcal{P}}^{\mathrm{quot}}(p, p') := \inf \{ d_{\mathcal{Q}}(q, q') : \varphi(q) = p, \, \varphi(q') = p' \}.
\end{equation}
This construction requires no structure on $\mathcal{P}$ beyond being the image of $\varphi$, and is therefore available under the central technical assumption (Assumption \ref{as: core}). The resulting $d_{\mathcal{P}}^{\mathrm{quot}}$ is a pseudo-metric satisfying $d_{\mathcal{P}}^{\mathrm{quot}}(\varphi(q), \varphi(q')) \leq d_{\mathcal{Q}}(q, q')$ by construction. This suffices for mathematical characterisation of phenomenological differences.

With this construction we can define the information length of phenomenological trajectories. Given a trajectory of beliefs $t \mapsto q_{\mu_t}$ for $t \in [0,1]$ the information length for phenomenology deriving from the quotient geometry is defined as
\begin{equation}
  \ell_{\mathcal{P}}^{\mathrm{quot}} = \int_0^1 d_{\mathcal{P}}^{\mathrm{quot}}(\varphi(q_{\mu_t}), \varphi(q_{\mu_t +d\mu_t})) \leq \int_0^1 d_{\mathcal{Q}}(q_{\mu_t}, q_{\mu_t +d\mu_t})= \ell_{\mathcal{Q}},
\end{equation}
and it is always bounded above by the Fisher information length of beliefs.

However, the quotient geometry does not retain the rich properties of the Fisher information geometry as it is not a Riemannian geometry. Only when the mapping from beliefs to phenomenology meets some regularity properties can we carry the Fisher geometry onto phenomenological space. For these reasons, it is the induced Fisher metric that we prefer when $\varphi$ meets those regularity conditions.

\subsection{Induced Fisher Geometry and Data Processing Inequality}
\label{app: induced fisher}

When $\mathcal{P}$ is a space of probability distributions and $\varphi$ arises from a measurable map between underlying sample spaces, then $\mathcal{P}$ inherits Fisher information structure from $\mathcal{Q}$. More precisely, let $\mathcal{Q}$ and $\mathcal{P}$ be spaces of distributions on sample spaces $X$ and $Y$ respectively, and let $\psi \colon X \to Y$ be a measurable map. The \emph{pushforward} $\varphi := \psi_\#$ maps each distribution $q \in \mathcal{Q}$ to a distribution $\varphi(q) \in \mathcal{P}$ by transforming the underlying sample space. Examples \ref{eg: nature}.1--3 of the central technical assumption (\cref{as: core}) all have this form: identity maps, marginalisations, and coarse-grainings are all pushforwards. For \cref{eg: nature}.4, where $\mathcal{P}$ need not be a space of distributions, Fisher geometry is unavailable and only quotient geometry applies. Under pushforward, the Fisher geometry can be induced on the phenomenological space and the data processing inequality provides substantive bounds on how information-geometric quantities transform.

\begin{proposition}[Induced Fisher geometry and data processing inequality]
    \label{prop: dpi length}
    Let $X$ and $Y$ be sample spaces, let $\mathcal{Q} = \{q_\mu\}$ be a parametric family of distributions on $X$, and let $\mathcal{P}$ be a space of distributions on $Y$. Let $\psi \colon X \to Y$ be a measurable map inducing $\varphi = \psi_\# \colon \mathcal{Q} \to \mathcal{P}$ via pushforward. Suppose $\varphi$ is sufficiently regular so that the Fisher information matrix
    \begin{equation}
    \label{eq: induced metric}
        g_{\mathcal{P}} := \left. \nabla_{d\mu}^2 \dkl[\varphi(q_\mu) \mid \varphi(q_{\mu + d\mu})] \right|_{d\mu = 0}
    \end{equation}
    is well-defined (see \cref{rem: reg}.\ref{rem: reg 1}). Then $\varphi$ induces information-geometric structures on $\mathcal{P}$: an information metric $g_{\mathcal{P}}$, information lengths $\ell_{\mathcal{P}}$ for trajectories $t \mapsto \varphi(q_{\mu_t})$, and a distance $d_{\mathcal{P}}$. They satisfy the following data processing inequalities:
    \begin{enumerate}[noitemsep]
        \item $g_{\mathcal{P}} \preceq g_{\mathcal{Q}}$ in the positive semi-definite ordering, where $g_{\mathcal{Q}}$ is the Fisher information metric in \eqref{eq: taylor}.
        \item For any trajectory $t \mapsto q_{\mu_t}$ in $\mathcal{Q}$, the information lengths satisfy $\ell_{\mathcal{P}} \leq \ell_{\mathcal{Q}}$.
        \item For any two beliefs $q_{\mu}, q_{\mu'} \in \mathcal{Q}$, the information distances satisfy $d_{\mathcal{P}}(\varphi(q_{\mu}), \varphi(q_{\mu'})) \leq d_{\mathcal{Q}}(q_{\mu}, q_{\mu'})$.
    \end{enumerate}
    Note that for some mappings $\varphi$, the Fisher information matrix can be singular in which case the information length and distances may be degenerate i.e. pseudo-metrics (see \cref{rem: reg}.\ref{rem: reg 2}).
\end{proposition}

\begin{remark}
    \label{rem: reg}
    \begin{enumerate}[noitemsep]
        \item \label{rem: reg 1} The regularity condition on $\varphi$ is automatically satisfied for \cref{eg: nature}.1--2: identity and projection maps are sufficiently regular. When $\varphi$ is a generic pushforward map (\cref{eg: nature}.3), the regularity condition for \eqref{eq: induced metric} intuitively constrains it to be second-order differentiable in $\mu$; standard coarse-grainings of common families of probability distributions satisfy this condition.
        \item \label{rem: reg 2} The induced metric $g_{\mathcal{P}}$ may be singular when $\varphi$ collapses distinct beliefs onto the same phenomenological state---that is, when $\varphi(q_\mu) = \varphi(q_{\mu'})$ for $\mu \neq \mu'$. In this case, $d_{\mathcal{P}}$ is a pseudo-metric rather than a metric capturing the idea that phenomenologically states that are indistinguishable would have zero distance regardless of underlying belief differences. This would apply if phenomenology were be a coarse-graining or a subset of beliefs.
        \item \label{rem: reg 3} When both quotient and Fisher geometries are available (\cref{eg: nature}.1--3), the two constructions need not coincide.
    \end{enumerate}
\end{remark}

\begin{proof}[Proof of \cref{prop: dpi length}]
    The data processing inequality for KL divergence states that for any measurable $\psi$,
    \begin{equation}
        \dkl[\varphi(q_\mu) \mid \varphi(q_{\mu'})] = \dkl[\psi_\# q_\mu \mid \psi_\# q_{\mu'}] \leq \dkl[q_\mu \mid q_{\mu'}].
    \end{equation}
    Substituting $\mu' = \mu + d\mu$ and expanding both sides to second order as in \eqref{eq: taylor}, the regularity assumption ensures the left-hand side admits the expansion $\frac{1}{2} d\mu \cdot g_{\mathcal{P}}(\mu) \, d\mu + o(\|d\mu\|^2)$, yielding
    \begin{equation}
        d\mu \cdot g_{\mathcal{P}}(\mu) \, d\mu \leq d\mu \cdot g_{\mathcal{Q}}(\mu) \, d\mu.
    \end{equation}
    where $g_{\mathcal{Q}}$ is the Fisher information metric on belief space \eqref{eq: taylor}. Since this holds for all $d\mu$, we have $g_{\mathcal{P}} \preceq g_{\mathcal{Q}}$, establishing (1).

    For (2), the metric inequality implies
    \begin{equation}
        \ell_{\mathcal{P}} = \int_0^1 \sqrt{d\mu_t \cdot g_{\mathcal{P}} \, d\mu_t}  \leq \int_0^1 \sqrt{d\mu_t \cdot g_{\mathcal{Q}} \, d\mu_t}  = \ell_{\mathcal{Q}}.
    \end{equation}

    For (3), the Fisher distance is the infimum of information length over paths. Since $\ell_{\mathcal{P}} \leq \ell_{\mathcal{Q}}$ for every path, the inequality is preserved under infima.
\end{proof}

\printbibliography

\end{document}